\documentclass[twocolumn,aps,prd,superscriptaddress,amsmath,
amssymb,showkeys,showpacs,nofootinbib] {revtex4-1}
\usepackage[utf8x]{inputenc}
\usepackage{graphicx}
\usepackage{bm}
\usepackage{hyperref}
\usepackage{dcolumn}
\usepackage{slashed}

\newcommand{\ev} {\ensuremath{\;\mathrm{eV}}}
\newcommand{\gev} {\ensuremath{\;\mathrm{GeV}}}
\newcommand{\be}{\begin{equation}}
\newcommand{\ee}{\end{equation}}
\newcommand{\ba}{\begin{eqnarray}}
\newcommand{\ea}{\end{eqnarray}} 
\def\lsi{\raise0.3ex\hbox{$<$\kern-0.75em\raise-1.1ex\hbox{$\sim$}}}
\def\gsi{\raise0.3ex\hbox{$>$\kern-0.75em\raise-1.1ex\hbox{$\sim$}}}
\begin{document}
\title{Resonant production of the sterile neutrino dark matter
and fine-tunings in the $\nu$MSM}
\author{Ananda Roy}
\email{ananda.roy@epfl.ch}
\affiliation{Indian Institute of Technology Kanpur, Kanpur, Uttar
Pradesh 208016,India}
\affiliation{Institut de Théorie des Phénomènes Physiques, EPFL,
CH-1015 Lausanne, Switzerland}
\author{Mikhail Shaposhnikov}
\email{mikhail.shaposhnikov@epfl.ch}
\affiliation{Institut de Théorie des Phénomènes Physiques, EPFL,
CH-1015 Lausanne, Switzerland}

\begin{abstract}
The generation of lepton asymmetry below the electroweak scale has a
considerable impact on production of dark matter sterile neutrinos.
Oscillations or decays of the heavier sterile neutrinos in the
neutrino minimal standard model can give rise to the
requisite lepton asymmetry, provided the masses of the heavier
neutrinos are sufficiently degenerate. We study the renormalization
group evolution of the mass difference of these singlet fermions
to understand the degree of necessary fine-tuning. We construct an
example of the model that can lead to a technically natural
realization of this low-energy degeneracy. 
\end{abstract}



\maketitle

\section{Introduction}
\label{sec1}
The neutrino minimal standard model ($\nu$MSM) is a renormalizable
extension of the standard model (SM) by three singlet Majorana
fermions with masses below the electroweak scale. As has been
elaborated in Refs. 
\cite{Asaka:2005an,Asaka:2005pn,Shaposhnikov:2008pf,Bezrukov:2007ep}
(for a review see \cite{Boyarsky:2009ix}), this model enables us to
solve, by economic means, four observational problems of the SM.
Owing to the low-energy see-saw mechanism the $\nu$MSM leads to
nonzero masses of the active neutrino flavors and thus to neutrino
oscillations. The model  provides a candidate for dark matter (DM)
particle in form of a sterile neutrino
\cite{Dodelson:1993je,Shi:1998km,Dolgov:2000ew, Abazajian:2001nj} in
the mass range of $1\ -\ 50$ keV (see \cite{Boyarsky:2009ix} for a
review). The coherent oscillations of the heavier neutral leptons
insure the
generation of baryon asymmetry of the Universe
\cite{Akhmedov:1998qx,Asaka:2005pn}. And lastly, a (large) nonminimal
coupling of the SM Higgs field with gravity leads to inflation
consistent with the cosmological observations \cite{Bezrukov:2007ep}. 

In the $\nu$MSM, the dark matter sterile neutrino is produced  at
temperatures $\sim 100$ MeV due to the mixing with active leptonic
flavors. The spectrum and the number density of produced particles
depends essentially on three parameters: the mixing angle with
ordinary neutrino, the sterile neutrino mass and the lepton asymmetry
of the Universe ($\Delta L$) at the production time (for a recent
analysis, see \cite{Asaka:2006nq,Laine:2008pg}; earlier considerations
can be found in \cite{Dodelson:1993je,Shi:1998km,Dolgov:2000ew,
Abazajian:2001nj}). The comparison of theoretical predictions with
cosmological and astrophysical observations (such as Ly-$\alpha$ data
and x-ray observations, constraining the free streaming length of DM
particles and their mixing angle with neutrinos, respectively) lead to
the conclusion that  the (low-temperature) lepton asymmetry must be
much larger than the baryon asymmetry $\Delta B \sim 10^{-10}$,
$\frac{\Delta L}{\Delta B}\geq 3\times 10^5$. \footnote{A possible way
to evade this requirement is some  modification of the $\nu$MSM,
allowing the DM sterile neutrino interactions with other new
particles, see, e.g.
\cite{Shaposhnikov:2006xi,Kusenko:2006rh,Anisimov:2008qs,Bezrukov:2009yw}.
Yet another possibility is related to primordial Higgs-inflation
\cite{Bezrukov:2008ut}.}

As was shown in \cite{Shaposhnikov:2008pf}, the presence of a pair of
nearly degenerate heavier neutral leptons in the $\nu$MSM may lead to
production of the requisite large lepton asymmetry below the
electroweak scale without a conflict with observed small baryon
asymmetry (generated by the same particles and by sphalerons at
electroweak temperatures). Basically, the resonant production of
$\Delta L$ occurs
at decoupling or during decays of singlet fermions which are taking
place well below the sphaleron freeze-out.

The requirement of generation of sufficient lepton asymmetry leads to
the stringent conditions on the parameters of the model, partially
analyzed in \cite{Shaposhnikov:2008pf}. The most important of them is
the level of the degeneracy of the pair of neutral leptons, which
demands a severe fine-tuning of the masses and couplings of the
$\nu$MSM. 

The aim of this paper is to study the stability of necessary
fine-tuning against radiative corrections (only the tree-level
analysis has been made in \cite{Shaposhnikov:2008pf}). In particular,
we will study the renormalization group (RG) evolution for the
mass difference of the singlet fermions and formulate the conditions 
that can lead to a technically natural realization of the low-energy
degeneracy, required for the low-temperature resonant leptogenesis,
essential for DM production. 

The paper is organized as follows: In Sec \ref{sec2} we review the
basic structure of the $\nu$MSM and its parametrization. In Sec
\ref{sec3} we explain the necessity of the degeneracy between neutral
leptons.  In Sec \ref{sec4} we describe and analyze the
renormalization group evolution of the essential parameters of the
model. In Sec \ref{sec5} we revisit different possible scenarios
for singlet fermion mass splitting and in Sec \ref{highdim} we
discuss an extention of the $\nu$MSM by higher-dimensional operators.
In Sec \ref{sec7} we discuss the phenomenological bounds
obtained as a result of the fine-tuning. Finally, in Sec
\ref{sec8}, we summarize our results. 

\section{The $\nu$MSM and its parametrization}
\label{sec2}
We use the Lagrangian of the $\nu$MSM in the following parametrization
\cite{Shaposhnikov:2008pf,Shaposhnikov:2006nn}:
\begin{equation}
 {\cal L}_{\nu MSM}={\cal L}_0+\Delta{\cal L}, 
\end{equation}
\begin{eqnarray*}
 {\cal L}_0={\cal L}_{SM}+
 \sum_{I=2,3}\overline{N_I}i\partial_\mu\gamma^\mu N_I
 - (h_{\alpha2}\overline{L_\alpha}N_2\tilde{\phi}
\end{eqnarray*}
\begin{equation}
 \hspace{20mm}+M\overline{N_2^c}N_3 +
h.c.),
\end{equation}
\begin{equation}
 \Delta{\cal L}=-h_{\alpha 3}\overline{L_\alpha}N_3\tilde{\phi}-\frac{\Delta
M}{2}\sum_{I=2,3}\overline{N_I^c} N_I + h.c.,
\end{equation}
where $N_I$ are the right handed singlet leptons ($I=2,3$), $\phi$ and
$L_\alpha(\alpha=e,\mu,\tau)$ are the Higgs and the lepton doublets,
respectively, $h$ is a matrix of the Yukawa coupling constants, $M$ is
the common mass of the two heavy neutral fermions, $\Delta M$ is the
diagonal element of the Majorana mass matrix, and
$\tilde{\phi}_i=\epsilon_{ij}\phi^*_{j}$, M and $\Delta M$ are taken
to be real.  We have omitted the DM sterile neutrino $N_1$ from the
Lagrangian as its influence on the problem we are interested in is
negligibly small (see \cite{Shaposhnikov:2008pf} for details). 

As has been shown in \cite{Shaposhnikov:2008pf}, one can solve for the
active neutrino masses explicitly:
\begin{equation}
m = \Big\{0,\frac{v^2}{M}[F_2F_3\pm|h^\dagger h|_{23}]\Big\}
\label{eq1}, 
\end{equation}
where $F_i^2\equiv[h^\dagger h]_{ii}$ and $v=174$ GeV is the vacuum
expectation value of the Higgs field. 
One encounters two different cases, namely the ``normal hierarchy,''
$m_1=0,~m_2=m_{sol}\simeq 0.009~\ev,~m_3=m_{atm}\simeq 0.05~\ev$, and
the ``inverted hierarchy,'' $m_1\approx m_2\approx m_{atm}\simeq
0.05~\ev,~m_3=0, |m_1 - m_2|\approx m_{sol}^2/(2m_{atm})\simeq 8\times
10^{-4}~\ev$. Normal hierarchy corresponds to the case when
$|h^\dagger h|_{23}\approx F_2F_3$, and the inverted hierarchy to the
case when $|h^\dagger h|_{23}\ll F_2F_3$. Here $m_{sol},~m_{atm}$ are
the solar and atmospheric neutrino mass differences (for a review see
\cite{Strumia:2006db}),
\ba
  \label{eq:sa}
  m^2_{sol} = 7.65^{+0.23}_{-0.20}\times 10^{-5} \ev^2,\\
  m^2_{atm} = 2.40^{+0.12}_{-0.11}\times 10^{-3} \ev^2.
\ea

For the lepton and baryon asymmetries of the Universe, one of the 
essential parameters is the ratio
\be
\epsilon^2\equiv \frac{(h^\dagger h)_{33}}{(h^\dagger h)_{22}}~,
\ee
measuring the relative strength of the coupling of the neutral leptons
$N_{2,3}$ to the active flavors. Without loss of generality it can
always be chosen in the region $\epsilon <1$.

The most important parameter is the mass difference between the two
heaviest neutrinos. Successful baryogenesis and leptogenesis
necessitates the mass difference to be small. At the tree level, there
are two contributions to the mass difference: one related to the
Majorana mass matrix and the other to the Higgs vacuum expectation
value and Yukawa couplings \cite{Shaposhnikov:2008pf}:
\begin{equation}
 \delta m_{tree}=\frac{|m^2|}{M},
 \label{trco}
\end{equation}
where
\begin{equation}
 m^2\equiv 2(h^\dagger h)_{23}v^2+ 2M \Delta M~.
\label{eq2}
\end{equation}
Note that $2|h^\dagger h|_{23}v^2 = M |\Delta m_\nu|$, where  $\Delta
m_\nu$ is the difference between active neutrino masses, and $\Delta
m_\nu\simeq 0.04~\ev~(8\times 10^{-4}~\ev)$ for normal (inverted)
hierarchy ($\Delta m_\nu$ by definition is a physical quantity and is
the RG invariant).

The analysis of radiative corrections to Eq. (\ref{eq2}) and its
RG evolution plays a central role in our considerations (see also
accompanying paper \cite{ananda1}).

\section{The fine-tunings necessary for low-temperature leptogenesis}
\label{sec3}
It has been shown in \cite{Shaposhnikov:2008pf} that the singlet
fermions  $N_{2,3}$ enter into thermal equilibrium at some temperature
$T_+ \sim M_W$ ($M_W$ being the intermediate vector boson mass) and
freeze out later at $T_- \gsi M$. Then they decay at $T_d \lsi M$. The
low-temperature lepton asymmetry can be generated at $T \sim T_-$ and
at  $T \sim T_d$, to satisfy Sakharov out of equilibrium conditions. 

The Yukawa coupling constants in the $\nu$MSM are very small,
\be
\label{F_2F_3}
F_2 F_3 = \frac{\sum m_\nu M}{v^2} \simeq 8
\times 10^{-16}\kappa\frac{M}{\gev}~,
\ee
where $\kappa=1$ ($\kappa=2$) for normal (inverted) hierarchy.
Therefore, the production of substantial lepton asymmetry can only be
possible in case of resonant $N_2 \leftrightarrow N_3$ transitions.
This means that the oscillation rate $\Gamma_{osc}$ should be of the
same order as the scattering rate
\be
\Gamma_s \simeq \frac{5 G_F^2 T^5 m_{atm}}{\epsilon M}~,
\ee
if the leptogenesis occurs at $T\sim T_-$, or the decay rate 
\be
\Gamma_N \simeq \frac{10 G_F^2 M^4 m_{atm}}{192\pi^3\epsilon }~,
\ee
if $T\sim T_d$. 

The quantity $\Gamma_{osc}$ is related to the difference of physical
masses of Majorana fermions $\delta m_{phys}$ as $\Gamma_{osc} \sim
\delta m_{phys}$, if $T<M$, or  $\Gamma_{osc} \sim M\delta m_{phys}/T$
for $T>M$. So, to get a substantial lepton asymmetry at $T\sim T_-$,
we have to require that
\begin{equation}\label{T-}
 \frac{M\delta m_{phys}}{T_{-}}=\frac{T_{-}^2}{M_0}~,
\end{equation}
where $M_0\approx M_{Pl}/1.66 \sqrt{g_{eff}},~M_{Pl}=1.2\times10^{19}$
GeV
and  the temperature dependence of the effective number of massless
degrees of freedom $g_{eff}$ may be taken from \cite{Asaka:2006nq}.
From (\ref{T-}), for $T\sim M\sim 1-10\ \text{GeV}$  we get, 
\begin{equation}
\label{decoup}
 \delta m_{phys}(T_-)~\lsi~ 10^{-18}-10^{-16}\ \text{GeV}~.
\end{equation}
An even stronger condition must be true for asymmetry generation in
$N$
decays. Taking again $M\sim 1-10$ GeV we have
\begin{equation}
\label{decay}
 \delta m_{phys}(T_d)~\lsi~ 10^{-23}-10^{-19}\ \text{GeV}~.
\end{equation}
The physical mass difference of singlet fermions in Eqs.
(\ref{decoup},\ref{decay}) should be taken at the corresponding
temperatures, $T_-$ or $T_d$.

It is the smallness of these numbers in comparison with the observed
neutrino mass difference $\Delta m_\nu$ (at least  $8\times 10^{-13}\
\text{GeV}$ for the inverted hierarchy case) that leads to the
fine-tuning problem. Indeed, there are several terms of different
nature which make up the physical mass difference: the tree level
Higgs contribution $\sim \Delta m_\nu$, the tree level Majorana
contribution $\sim \Delta M$, the zero-temperature loop corrections,
and finite-temperature corrections. The latter effects were analyzed
in detail in \cite{Shaposhnikov:2008pf}, Sec. 5.1.  It was found
there that these corrections  lead to the induced finite-temperature
mass difference $\delta m_{phys}$ which is of the order of the Hubble
rate $H$ at temperature $T_-$ (see Eq. (7.32) of
\cite{Shaposhnikov:2008pf}) or smaller than $H$ at $T_d$. In other
words, they do not spoil the resonant character of low-temperature
leptogenesis, provided the zero-temperature mass difference is tuned
to small values, given in Eqs. (\ref{decoup},\ref{decay}) above.
Therefore, we will concentrate on zero-temperature contributions in
what follows. A part of the zero-temperature loop corrections can be
absorbed into the tree-level Higgs contribution, converting it to the
physical mass difference of active neutrinos
\cite{Shaposhnikov:2008pf}:
\be
 \delta m_{phys}=|\Delta m_\nu e^{i\alpha} +\Delta M| + 
 {\cal O}(\frac{\Delta m_\nu}{16\pi^2}\frac{M^2}{v^2})~,
\label{ft}
\ee
where $\alpha$ is some phase. Numerically, the last term in (\ref{ft})
is of the order of $10^{-18}\ \text{GeV}$ (for $M\sim 1$ GeV), which
is of the same order as (\ref{decoup}) and is  much larger than
(\ref{decay}). Therefore, the tuning of the tree-level Majorana
contributions to the {\em physical value} $\Delta m_\nu$ must be done
together with radiative corrections in order to achieve the required
degeneracy $\delta m_{phys}\ll \Delta m_\nu$. 

Yet another attitude can be used in discussing the fine-tuning.
Suppose that the compensation of the {\em tree} contributions  in
(\ref{trco}) is associated with some symmetry which may potentially
exist at some high energy scale, such as the Planck mass $M_{Pl}$.
Then the physical mass-difference $\delta m_{phys}$, which is
determined by the {\em low-energy} parameters, is not in general zero,
due to the running of these parameters. The consideration of the
running of the relevant parameters, together with computation of the
radiative corrections \cite{ananda1}, would allow then to estimate the
"natural" values of the mass difference. This is the purpose of the
next section.

\section{RG evolution of the mass-difference}
\label{sec4}
The RG running of the Majorana masses and of sterile-active Yukawa
couplings can be extracted from \cite{Antusch200287,Lin:2009sq}
\be
\label{MR}
(4\pi)^2\frac{d}{dt}M_R=(h^\dagger h)M_R+M_R(h^\dagger h)^T~,
\ee
\begin{eqnarray*}
\label{rgh}
(4\pi)^2\frac{d}{dt}h=\Big\{\frac{3}{2}hh^\dagger-\frac{3}{2}
Y_e^\dagger Y_e-\frac{3}{4} g_1^2-\frac{9}{4} g_2^2
\end{eqnarray*}
\begin{equation}
\hspace{10mm} +\text{Tr}[3Y_u^\dagger Y_u+3Y_d^\dagger Y_d
+hh^\dagger+Y_e^\dagger Y_e]\Big\}h, 
\end{equation}
where $t\equiv\ln\frac{\mu}{\mu_0}$ (for definiteness we take $\mu_0=$
top quark mass),  $M_R$ is the Majorana mass matrix, $Y_e$ is the
diagonal charged lepton matrix, $Y_{u(d)}$ is the Yukawa coupling
matrix for the up (down) quarks, and $g_{1},~g_2$ represent the gauge
coupling for U(1) and SU(2), respectively. Since (\ref{eq2}) contains
the vacuum expectation value of the Higgs field, $v^2=m_H^2/2\lambda$
($m_H^2$ is the mass parameter in the SM, $\lambda$ is the scalar
self-coupling)\footnote{In this paper, we use
the Higgs potential form as:\\
$V(\phi)=-\mu^2(\phi^\dagger\phi)+\frac{\lambda}{2}
(\phi^\dagger\phi)^2$.}, we will also need the RG runnings of
$m_H^2$ and
$\lambda$. They are given, for example, in \cite{Davies1990431}:
\be
\label{m_H}
(4\pi)^2\frac{dm_H^2}{dt}=m_H^2\Big(6\lambda-\frac{3}{2}g_1^2-\frac{9}
{2}g_2^2+6y_t^2\Big),
\ee
\begin{eqnarray*}
\label{lambda}
(4\pi)^2\frac{d\lambda}{dt}
=(12\lambda-3g_1^2-9g_2^2)\lambda+\frac{3}{4}g_1^4
\end{eqnarray*}
\begin{equation}
\hspace{25mm}+\frac{3}{2}g_1^2g_2^2+\frac{9}{4}
g_2^4+12(\lambda-y_t^2)y_t^2 .
\end{equation}

To complete the system, we add the RG equation for the  top quark
Yukawa coupling $y_t$,
\be
(4\pi)^2\frac{dy_t}{dt}=y_t\Big(\frac{9}{2}y_t^2-\frac{17}{12}g_1^2-
\frac{9}{4}g_2^2-8g_3^2\Big).
\ee
The RG evolution of other parameters of the SM will not be needed in
this section.

Using the fact that  $h \ll g_{1,2}$, and neglecting numerically small
Yukawa couplings of charged leptons and quarks (except  $y_t$), we get
the RG equation for the Majorana mass difference and for $h$:
\begin{equation}\label{RGDeltaM}
 (4\pi)^2\frac{d}{dt}\Delta M=2M|h^\dagger h|_{23}~,
\end{equation}
\begin{equation}\label{yukawacoup}
 (4\pi)^2\frac{d}{dt}h=h\Big\{3y_t^2-\frac{3}{4}g_1^2-\frac{9}
{4}g_2^2\Big\}.
\end{equation}
We consider the Yukawa couplings for the sterile neutrinos to be real
in this section. This simplifies the computations, but does not change
the numerical estimates and qualitative conclusions.
 
So, the change in $\Delta M$ due to RG evolution from the Planck scale
to $\mu_0$ is of the order of 
\be
\label{large}
\frac{\Delta
m_\nu}{(4\pi)^2}\ln\left(\frac{M_{Pl}}{\mu_0}\right)\frac{M^2}{v^2}
\simeq 
(6.5 \times10^{-18}-4\times 10^{-16})\frac{M^2}{\text{GeV}}~,
\ee
depending on the type of neutrino hierarchy. This number for the
inverted hierarchy is roughly of the same order as (\ref{decoup}), but
exceeds considerably (\ref{decay}). 

The RG evolution of the Higgs contribution to the mass difference is
much more substantial. In order to make an  estimate of this
contribution, we fix the initial condition for the Yukawa couplings at
the Planck scale and run them down to the electroweak scale. For
definiteness we take the inverted hierarchy case, the conclusions for
the normal hierarchy are the same.

It was shown in \cite{Shaposhnikov:2008pf} that for inverted hierarchy
the coupling of $N_2$ to $\tau$ flavour (consideration of other
leptonic families do not change the results) is
\begin{equation}
 |h_{\tau
2}|=\frac{F_2}{2}\Big|\cos\theta_{12}e^{-i\zeta}+i\sin\theta_{12}e^{
i\zeta}\Big|,
\end{equation}
where $\tan^2\theta_{12}\simeq 0.48$ from active neutrino oscillation
data \cite{Strumia:2006db} and $\zeta$ is a phase factor that varies
between $0$ and $2\pi$. A little algebra shows that
\begin{equation}\label{maxmin}
 \Big|\frac{h_{\tau 2}}{F_2}\Big|_{min}=0.14,\ \Big|\frac{h_{\tau
2}}{F_2}\Big|_{max}=0.7~.
\end{equation}
Similar relations hold for $|h_{\tau 3}|$ with $F_2$ replaced by
$F_3$.

Using \eqref{maxmin} and \eqref{F_2F_3}, it follows that the
maximum and minimum values of the Yukawa couplings, for $M=1$ GeV and
$\epsilon =1$, are 
\begin{equation}
\label{maxmin1}
 |h_{\tau 2}|_{min}\sim 3.4\times10^{-9},\ |h_{\tau 2}|_{max}\sim
1.7\times10^{-8}~.
\end{equation}
A similar set of maximum and minimum values hold for $|h_{\tau 3}|$.
The
values of Yukawa couplings scale as $h_{\alpha 2} \propto
\sqrt{M/\epsilon}$, $h_{\alpha 3} \propto
\sqrt{M\epsilon}$.

\begin{figure}[htp]
 \centering
\includegraphics[height=40mm,width=80mm]{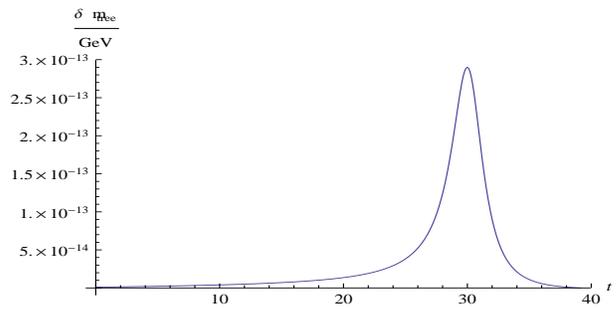}
\caption{RG evolution of $\delta m_{tree}$ for a Higgs mass of $\simeq
145\ GeV$.}
\label{deltamtree}
\end{figure}

Using the above-mentioned bounds along with Eqs.
\eqref{m_H},\eqref{lambda} and \eqref{yukawacoup}, we obtain the RG
evolution of the Higgs contribution down to the electroweak scale. 
Performing the necessary numerical computation, we find that the Higgs
contribution to the mass-difference ranges from $10^{-10}$ GeV to
$10^{-11}$ GeV depending on whether the maximum or minimum
value is chosen for the Yukawa couplings. This value is of the order
of $\Delta m_\nu$, meaning that if  $m^2$ in (\ref{eq2}) is tuned to
zero (implying $\delta m_{tree}=0$) at the Planck scale, the
physical mass difference between $N_2$
and $N_3$ will generically be of the order of $\Delta m_\nu$,
exceeding considerably the required values (\ref{decoup},\ref{decay}).
A way out is a fine-tuning of the Higgs mass, which can be chosen in
such a way that $\delta m_{tree}(M_{pl})=\delta m_{tree}(M_{W})=0$. In
Fig.
\ref{deltamtree}, we give the RG evolution of the mass-difference
$\delta m_{tree}(\mu)$ for a particular choice of Higgs mass (145
GeV), which leads to this situation. 
%%%%%%%%%%%%%%%%%%%%%%%%%%%%%%%%%%%%%%%%%%%%%%%%%%%
%%%%%%%%%%%%%%%%%%%%%%%%%%%%%%%%%%%%%%%%%%%%%%%%%%%

\section{Scenarios for singlet fermion mass difference revisited}
\label{sec5}
To summarize, if there were a compensation in the tree-level
contribution to the mass-difference $\delta m_{tree}$ due to existence
of some symmetry at a higher-energy scale $(M_{pl})$, it  does not
exist at
the electroweak scale and the physical mass difference acquires a
nonzero value, which is of the same order as that of the change in
Higgs contribution: $10^{-10} - 10^{-11}$ GeV. The reason is that the
Majorana contribution to $m^2$ runs much more slowly than the Higgs
one.
One can arrive to (\ref{decoup},\ref{decay}) either by tuning of the
Higgs mass, or by tuning the initial condition for $\Delta M$ at the
Planck scale. In Sec. \ref{highdim} we will discuss how this
situation can be changed in some extension of the $\nu$MSM. Meanwhile,
we will describe the possible scenarios (sf.
\cite{Shaposhnikov:2008pf}) for singlet fermion mass difference
accounting for RG behavior studied in Sec. \ref{sec4}. Depending on
the relative importance of the different contributions to $\delta
m_{phys}$, they may be classified as follows:

\begin{itemize}
 \item \textbf{Scenario Ia}: $\Delta M(M_{pl})=0$. In this case the
physical  mass difference is mostly due to the tree-level Higgs
condensate and loop corrections. One can easily check from Eqs.
\eqref{eq1},\ \eqref{eq2} that this leads to 
\begin{equation}
 \delta m_{phys}\approx \Delta m_\nu.
\end{equation}
Thus, $\delta m_{phys}\simeq m_{atm}-m_{sol}\approx 5\times10^{-11}$
GeV for normal hierarchy and $\delta m_{phys}\simeq\frac{\Delta
m_{sol}^2}{2m_{atm}}\approx8\times10^{-13}$ GeV for inverted. 

\item \textbf{Scenario Ib}: This corresponds to the situation in which
the tree-level Higgs and the Majorana contribution are of the same
order of magnitude at the Planck scale, including the case when there
is a compensation of the two contributions, $m^2(M_{pl})=0$. In this
case and without any special fine-tuning one gets
\begin{equation}
 \delta m_{phys}\sim \Delta m_\nu.
\end{equation}

\item \textbf{Scenario II}: The {\em physical} mass difference of the
singlet fermions is much smaller than the active neutrino
mass difference, i.e.
\begin{equation}
 \delta m_{phys}\ll\Delta m_\nu.
\end{equation}
Only this scenario can lead to production of substantial
low-temperature lepton asymmetry and thus to resonant sterile neutrino
dark matter production. It requires a fine-tuning between
contributions of the different nature (Higgs, Majorana, and loop
corrections) and thus is ``unnatural" in a technical sense
\footnote{By
``technically" natural we mean the situation in which the fine-tuning
made at the high energy scale persists to small energies.}.

\item \textbf{Scenario III}: The Majorana contribution dominates,
in which case 
\begin{equation}
 \delta m_{phys}\gg \Delta m_\nu.
\end{equation}
\end{itemize}

The scenarios \textbf{Ia, Ib}, and \textbf{III} are natural in
the technical sense. However, they do not lead to low-temperature
lepton
asymmetry. In Sec. \ref{highdim} we will present an extension of the
$\nu$MSM in which \textbf{Scenario II} can be realized as a natural
possibility.
%%%%%%%%%%%%%%%%%%%%%%%%%%%%%%%%%%%%%%%%%%%%%%%%%%%%%%%%
\section{Higher-dimensional operators}
\label{highdim}
The $\nu$MSM can be extended in several ways, which may break the 
relations (\ref{eq1},\ref{trco},\ref{eq2}), being the basis of the
analysis of the previous sections. The simplest possibility, in the
spirit of effective field theories, is to add higher-dimensional
operators. There are five independent five-dimensional operators which
can be constructed from the fields of the $\nu$MSM. One contains the
fields of the SM only,
\be
c_1^{\alpha\beta}\overline{L_\alpha}\tilde{\phi}\phi^\dagger L^c_\beta~.
\label{LL}
\ee
The other two include the singlet fermions as follows:
\begin{equation}
c_2^{IJ}\overline{N^c_I} N_J \phi^\dagger\phi,~
c_3^{IJ}\overline{(\partial_\mu N_I)^c}\partial_\mu N_J.~
\label{NN}
\end{equation}
Yet another two include $L$ and $N$ simultaneously,
\begin{equation}
c_4^{\alpha I}\overline{\slashed{D}L_\alpha}N^c_I\tilde{\phi},~
c_5^{\alpha I}\overline{L_\alpha}(\slashed{D}N_I)^c\tilde{\phi}.
\label{NL}
\end{equation}
Here $c_i$ are new coupling constants with dimension GeV$^{-1}$.
The operator (\ref{LL}) can change the relation (\ref{eq1}), whereas
operators (\ref{NN},\ref{NL}) can change (\ref{trco},\ref{eq2}).

There are several possible ways how the \textbf{Scenario II} can be
made technically natural. The first one is based on the use of the
operator (\ref{LL}), which contributes to the active fermion mass
difference. The idea is as follows. Suppose that at $M_{pl}$ the
following two relations hold simultaneously: 
\be
\Delta M=0~,~~~~(h^\dagger h)_{23}=0~. 
\label{plcond}
\ee
These initial conditions require that the active neutrino mass
hierarchy must be inverted, as follows from Eq. \eqref{eq1}.
Then, if charged lepton Yukawa couplings $Y_e$ are set to zero, the
Lagrangian of the $\nu$MSM has an extra global leptonic U(1) symmetry
\cite{Shaposhnikov:2006nn}. This means, that the relations
(\ref{plcond}) are RG scale-independent, and that the singlet fermions
will remain exactly degenerate.  In fact, the charged leptonic Yukawa
couplings violate explicitly this symmetry and thus lead to the
breaking of exact degeneracy at small energies. So, some mass
difference between the singlet fermions will be generated. As we will
see below, it is much smaller than the observed mass difference
between active neutrinos.  The inclusion of the operator (\ref{LL}) is
needed  to generate the observed active neutrino mass difference and
requires $c_1 \simeq 1/(10^{16}{\mbox GeV})$; the common mass of
active neutrinos is due to $N_{2,3}$.

Let us make an estimate of the splitting between $N_{2,3}$ owing to
charged lepton Yukawas. Among $Y_e$ the tau-lepton coupling is the
largest, and  $Y_e^\dagger Y_e$, which was previously omitted from Eq.
(\ref{rgh}), has the form
\begin{equation}
Y_e^\dagger Y_e=
\begin{pmatrix}
0&0&0\\
0&0&0\\
0&0&y_\tau^2
\end{pmatrix}.
\end{equation}

\begin{figure}[htp]
 \centering
\includegraphics[height=40mm,width=80mm]{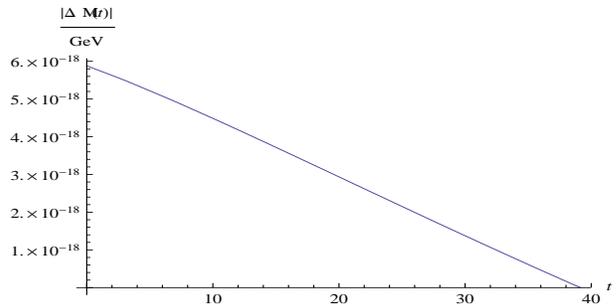}
\caption{RG evolution of $\Delta M$.}
\label{DeltaM}
\end{figure}

\begin{figure}[htp]
 \centering
\includegraphics[height=40mm,width=80mm]{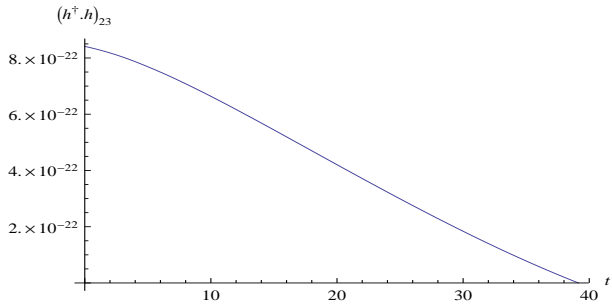}
\caption{RG evolution of $|h^\dagger h|_{23}$.}
\label{hhdag}
\end{figure}

The evolution of $\Delta M$ and $(h^\dagger h)_{23}$, following from
Eqs. (\ref{MR},\ref{rgh}) with initial conditions (\ref{plcond}) is
shown in Figs. \ref{DeltaM} and \ref{hhdag} respectively.  We
give only the plots corresponding to the minimum value of the singlet
fermion Yukawa couplings [see \eqref{maxmin1}] and for $M=1$ GeV. 
%%%%%%%%%%%%%%%%%%%%%%%%%%%%%%%%%%%%%%%%%%%%%%%%
%%%%%%%%%%%%%%%%%%%%%%%%%%%%%%%%%%%%%%%%%%%%%%%%
The energy scale has been varied from the top quark mass ($t=0$) to
Planck scale ($t=39.14$). We can see that both $|\Delta M|$ and 
$(h^\dagger h)_{23}$ decrease steadily with the increase of $t$.
%%%%%%%%%%%%%%%%%%%%%%%%%%%%%%%%%%%%%%%%%%%%%%%%%
%%%%%%%%%%%%%%%%%%%%%%%%%%%%%%%%%%%%%%%%%%%%%%%%%

Numerically, for the minimal value of $h_{\tau 2}$ [\textbf{case (a)}
for future reference], the mass
difference is given by (sf eq. (\ref{ft})
\be\label{treemassa}
\delta m_{phys}\simeq\Big|5\times 10^{-17} e^{i\alpha} 
- 1.2\times 10^{-17} \frac{M^2}{{\rm
GeV}^2}\Big|~{\rm GeV}~,
\ee
establishing a typical scale $\sim 10^{-17}$ GeV.

Analogous computation for the maximal value of $h_{\tau 2}$
[\textbf{case (b)}] leads us to the similar dependence: the
mass-difference is given by
\begin{equation}
\label{treemassb}
\delta m_{phys}\simeq \Big|1.3\times10^{-15}e^{i\alpha}-
3\times10^{-16}\frac{M^2}{\text{
GeV}^2} \Big|~\text{GeV},
\end{equation}
with a typical scale of $\sim 10^{-16}$ GeV. 

In order to find the different possible values of mass-difference, we
analyze the two cases separately for different values of M. 
\begin{itemize}
 \item \textbf{case (a):} From Eq. \eqref{treemassa} we see that for a
particular choice of the Majorana mass ($M\simeq2$ GeV), the
mass difference can be made arbitrarily small by a suitable choice of
the Majorana phase ($\alpha$). In particular, the mass difference
vanishes for $\alpha=0$. Therefore, it may be safely concluded that
for $M\simeq2$ GeV, the following bounds hold true for $\delta
m_{phys}$: 
\begin{equation}\label{bounda}
 0\leq\delta m_{phys}\leq 10^{-16} {\rm\ GeV}.
\end{equation}
We now turn to the two other possibilities: $M\ll2$ GeV and $M\gg2$
GeV. Considering $M\ll2$ GeV first, we can see that
\begin{equation}\label{Mll2a}
 \delta m_{phys}=5\times10^{-17}{\rm\ GeV},
\end{equation}
while for $M\gg2$ GeV, 
\begin{equation}\label{Mgg2a}
 \delta m_{phys}=1.2\times10^{-17}\frac{M^2}{\rm GeV}.
\end{equation}

\item \textbf{Case (b):} Performing similar analysis, we see that once
again for $M\simeq2$ GeV, the mass difference becomes arbitrarily
small, depending on the choice of the Majorana phase. We thus
establish the bound for $\delta m_{phys}$ in this case: 
\begin{equation}\label{boundb}
 0\leq\delta m_{phys}\leq 2.6\times10^{-15} {\rm\ GeV}.
\end{equation}
For $M\ll2$ GeV, 
\begin{equation}\label{Mll2b}
 \delta m_{phys}=1.3\times10^{-15}{\rm\ GeV},
\end{equation}
while for $M\gg2$ GeV, 
\begin{equation}\label{Mgg2b}
 \delta m_{phys}=3\times10^{-16}\frac{M^2}{\rm GeV}.
\end{equation}

\end{itemize}

Thus, we see that the mass difference $\delta m_{phys}$ so obtained in
the two cases is indeed much smaller than the active neutrino
mass difference, what is needed for the low-temperature resonant
leptogenesis. As we have already said, the role of the
higher-dimensional operator (\ref{LL}) is to provide the additional
contribution to mass difference of the active flavors. The amplitude
of all other operators must be small enough in order not to spoil the
relations (\ref{treemassa},\ref{treemassb}), which is technically
natural. In Sec. \ref{sec7} we will consider different bounds
on parameters of the $\nu$MSM which appear in this scenario.

Yet another possible way to make the \textbf{Scenario II}
technically natural is based on operators (\ref{NL}).  These
operators contribute to the singlet fermion mass  with the terms of
the order  $c_4^{\alpha I} h_{\alpha J} v^2,~c_3^{\alpha I} h_{\alpha
J} v^2$. If the RG running of some combination of  these
contributions happens to be the same as the running of the Higgs
contribution $2(h^\dagger h)_{23}v^2/M$, this can be used to cancel
the largest effect (\ref{large}). If true, the physical mass
difference can be ``naturally" made of the order of the last term in
Eq. (\ref{ft}). The analysis of this possibility goes beyond the scope
of the present paper.  

%%%%%%%%%%%%%%%%%%%%%%%%%%%%%%%%%%%%%%%%%%%%%%%%%%%%%%%%%%%%%%%%%%%%%
\section{Phenomenological bounds on parameters of
$\nu$MSM from naturalness}
\label{sec7}
Let us assume that the physical mass difference is indeed given by the
relations derived in Sec. \ref{highdim}. Is the baryogenesis due to
singlet fermion oscillations still operational? What can be said 
about the parameters $\epsilon$ and $M$ from the requirement of
resonant dark matter production? 

To answer the first question, let us determine the crucial parameter of
the baryogenesis -- the number of oscillations $x(T_{\rm sph})$ of
singlet fermions before the freezing of sphalerons at temperature
$T_{\rm sph} \sim 150$ GeV \cite{Shaposhnikov:2008pf}. The
high-temperature mass difference of the singlet fermions comes from
two-loop graphs  which include the square of charged lepton Yukawa
coupling and square of $h_{\alpha I}$ and corresponds to the double
Higgs exchange. It can be estimated as 
\be 
\Delta M(T)^2 \sim
\left(\frac{h_{\tau 2}y_\tau}{16}\right)^2 T^2~,
\label{dd}
\ee 
leading to 
\be 
x(T_{\rm sph}) \simeq \left(\frac{h_{\tau 2}y_\tau}{16}\right)^2
\frac{M_0}{T_{\rm sph}}~.
\ee
Yet another contribution to $\Delta M(T)^2$ comes from the Higgs
condensate. It is of the same order of magnitude as (\ref{dd}), since
at the sphaleron freeze out $v(T) \sim T$. Numerically, $x(T_{\rm
sph})\sim (10^{-7}-10^{-6})/\epsilon$. Since baryon asymmetry is
proportional to $x$ for $x \ll 1$, it is automatically smaller than
the low-temperature lepton asymmetry by a factor $\sim 10^6$ (for
$\epsilon \sim 1$), providing a potential explanation of this
hierarchy. At the same time, with this value of $x$, the baryon
asymmetry can be as large as $\Delta B\sim 10^{-9}$, exceeding the
observed one. In other words, the answer to the first question is
positive.

To analyze the second question, we will consider two choices of
$h_{\tau 2}$, referred to as (\textbf{case (a)} and \textbf{case (b)})
above. As has been shown in \cite{Shaposhnikov:2008pf}, the lepton
asymmetry can be generated in either by decays or oscillations of the
sterile neutrinos. We will consider these two situations separately.

\subsection{Leptogenesis by decay of sterile neutrinos}
\label{dec}
 Considering the case of decays first, it has been shown in 
\cite{Shaposhnikov:2008pf} that the maximal lepton asymmetry  $\Delta$
which can be generated in decays of $N_{2,3}$ is of the order 
\begin{equation}\label{lept}
 \Delta\approx\Delta _{max}\frac{\epsilon M^2}{M_0\delta
m_{phys}}~,
\end{equation}
where $\Delta_{\max}=2/11$. The condition that the decays of $N_{2,3}$
occur above the dark matter creation temperature $\sim 100$ MeV reads
\begin{equation} \label{M-lower bound}
\frac{M}{\text{GeV}}>1.4\Big(\frac{\epsilon}{2\times10^{-3}}\Big)   
^{\frac{1}{4}}. 
    \end{equation}
To produce a necessary amount of dark matter sterile neutrinos, it is
required that \cite{Laine:2008pg}
\be 
\label{max}
\Delta\geq2\times10^{-3}~. 
\ee

\begin{itemize}
 \item \textbf{Case (a):} When $M\ll2$ GeV, using
\eqref{Mll2a},\eqref{lept} and \eqref{M-lower bound}, we arrive at the
minimum value of M to be $3$ GeV, which contradicts the starting
assumption that $M\ll2$ GeV. On the other hand, for $M\gg2$ GeV, using
\eqref{Mgg2a} and \eqref{lept}, we arrive at a lower bound on
$\epsilon$:
\begin{equation}
 \epsilon \gsi 0.1.
\end{equation}
Corresponding minimum value of M [from \eqref{M-lower bound}] is
$\sim4$ GeV. 

\item \textbf{Case (b):} Performing a similar analysis as before, we
once again reject the case when $M\ll2$ GeV. For the case when $M\gg2$
GeV, we obtain a lower bound on $\epsilon$ to be $2.3$, conflicting
with the condition $\epsilon <1$.  
\end{itemize}

To summarize, a sufficient lepton asymmetry in decays of  $N_{2,3}$
can be generated in \textbf{case (a)} for $M\gsi 4$ GeV and $\epsilon
\gsi 0.1$. Or, one has to require the fine-tuning $M\simeq 2$ GeV and
$\alpha\ll 1$, making the physical mass difference even smaller than
$\sim 10^{-16}$ GeV.

\subsection{Leptogenesis from coherent oscillations of sterile
neutrinos}
\label{osc}
As has been shown in \cite{Shaposhnikov:2008pf}, the most important
parameter which determines the value of lepton asymmetry generated  at
temperature $T_-$ is the number of oscillations  $x(T_-)/2\pi$ of 
$N_{2,3}$ given by
\begin{equation}\label{x}
 x(T_-)=\frac{0.15\kappa B}{\epsilon}(G_FM_0)^2m_{atm}
 \delta m_{phys}(T)~,
\end{equation}
where $\kappa$ is 1 or 2 depending on normal or inverted hierarchy,
$B=5$ and $G_F$ is the Fermi coupling
constant. The case $\delta m_{phys}(0)=0$, in which $\delta
m_{phys}(T_-)\neq 0$ due to finite-temperature effects, leads to 
$x(T_-)\equiv x_T\simeq 10$ and to some region in the $\epsilon,M$
plane leading to the required lepton asymmetry (\ref{max}). This
analysis stays in force also if $\delta m_{phys}(0)\neq 0$, provided 
\be
\label{mmc}
\delta m_{phys}(0)<\frac{x_T \epsilon}{0.15\kappa B(G_FM_0)^2m_{atm}}~.
\ee
We will give below the corresponding parts of the phase space. For
this end we find  $x(T_-)$ from Eq. (\ref{x}) replacing $\delta
m_{phys}(T)$ by $\delta m_{phys}$ defined in Eqs.
(\ref{bounda}-\ref{Mgg2b}).

\begin{itemize}
 \item \textbf{Case (a)}: For $M\ll2$ GeV, using \eqref{Mll2a} and
\eqref{x}, we get
\begin{equation}
 x(T_-)=\frac{0.25}{\epsilon},
\end{equation}
which implies $\epsilon\geq 0.025$ to satisfy inequality $x(T_-)<
x_T$.  For $M\gg2$ GeV, using \eqref{Mgg2a} and
\eqref{x}, we get
\begin{equation}
 x(T_-)=\frac{0.06}{\epsilon}\frac{M^2}{\text{GeV}^2},
\end{equation}
which implies $\epsilon\geq0.006M^2/\text{GeV}^2$. Plotting
$\epsilon$ as a function of M, for the two situations, the
allowed values of $\epsilon-M$ are depicted by the shaded portion in
Fig. \ref{min_val_bounds_on_mass_a}. 
\begin{figure}[htp]
 \centering
\includegraphics[height=40mm,width=80mm]{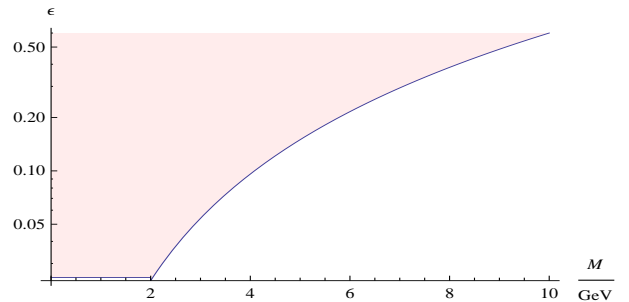}
\caption{Shaded region of the plot represents the values of
$\epsilon$-M in which the analysis of Ref. \cite{Shaposhnikov:2008pf}
is not changed for the \textbf{case (a)}. The $\epsilon$ axis is
shown in logarithmic scale. }
\label{min_val_bounds_on_mass_a}
\end{figure}

%%%%%%%%%%%%%%%%%%%%%%%%%%%%%%%%%%%%%%%%%%%%%%%%%%%%%%%%%%%%%%%%%%%%%
\item \textbf{Case (b)}: Similar analysis for  $M\ll2$ GeV leads to
\begin{equation}
 x(T_-)=\frac{6.4}{\epsilon},
\end{equation}
which implies $\epsilon\geq 0.64$.
Again for $m\gg2$ GeV, we arrive at the
relation:
\begin{equation}
 x(T_-)=\frac{1.48}{\epsilon}\frac{M^2}{\text{GeV}^2}.
\end{equation}
which implies $\epsilon\geq0.148M^2/\text{GeV}^2$. These regions are
shown in Fig. \ref{min_val_bounds_on_mass_b}. 
\begin{figure}[htp]
 \centering
\includegraphics[height=40mm,width=80mm]{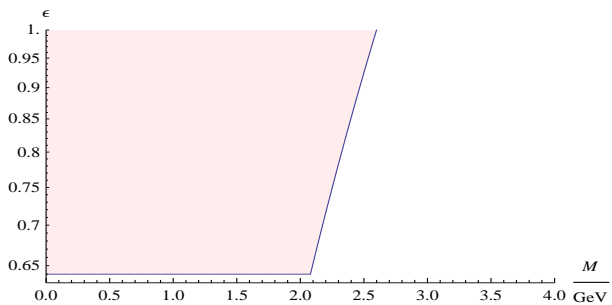}
\caption{The same as in Fig. 4 for the \textbf{case (b)}. In this
plot, we only show the region where $\epsilon\leq1$, which we assume
to be true throughout the paper. }
\label{min_val_bounds_on_mass_b}
\end{figure}
\end{itemize}

As in Sec. \ref{dec}, the case when $M\simeq 2$ GeV and $\alpha\ll 1$
is special, and all values for $\epsilon$ and $\delta m_{phys}$, found
in Sec. \ref{highdim} are allowed.

\section{Conclusions}\label{sec8}
The generation of lepton asymmetry below the sphaleron
freeze-out temperature enables generation of a large lepton asymmetry
without leaving a trace on the much smaller baryon asymmetry of the
Universe. The production of dark matter sterile neutrinos is dependent
on the lepton asymmetry present at the time of production. 
  In this paper we studied the RG evolution of the
mass splitting between neutral leptons $N_{2,3}$ of the $\nu$MSM,
essential for resonant production of low-temperature lepton asymmetry,
followed by resonant production of dark matter sterile neutrinos. We
found that the mass differences of the order or greater than the
observed mass differences in the active neutrino sector are natural
in the technical sense. In other words, the RG running of it from the
Planck to the low-energy scale leads to corrections $\sim \Delta
m_\nu$.

At the same time, the low-temperature resonant leptogenesis requires
the splitting that is much smaller than $\sim \Delta m_\nu$ and thus
is ``fine-tuned." We described an extension of the  $\nu$MSM by
higher-dimensional operators, in which this fine-tuning is due to an
approximate symmetry of the theory at the Planck scale and which is
not spoiled by the RG evolution. It requires the hierarchy of neutrino
masses to be inverted. We analyzed the constraints on the masses and
couplings of the singlet leptons in this scenario and demonstrated its
feasibility.

\begin{acknowledgments}
We thank Dmitry Gorbunov for discussions and helpful comments. 
This work was supported in part by the Swiss National Science
Foundation.  It was facilitated by the Student Exchange program
between Ecole Polytechnique Federale de Lausanne and Indian Institute
of Technology, Kanpur. 
\end{acknowledgments}

\bibliography{all}

\end{document}